\documentstyle[manuscript,prbbib,aps,epsfig,12pt,epsf,amstex,amsfonts,amssymb,pstricks]{revtex}
\numberwithin{equation}{section}

\date{version of \today}
\begin{document}
\bibliographystyle{pccp}

\title{Nearside-farside theory of differential cross sections : \newline resummation of a partial wave series involving Legendre polynomials}

\author{C. Noli and J. N. L. Connor}
\address{Department of Chemistry , University of Manchester, Manchester M13 9PL, England}

\maketitle

\newpage 
\begin{abstract}
$\:\:\:\:\:$ We report a new resummation procedure for the partial wave series (PWS) representation of the scattering amplitude, when a basis set of Legendre polynomials is used for the expansion. The effect of the resummation is to remove from the PWS the factor $( \alpha + \beta \cos \theta) ^{-r}$ where $\theta\:$ is the scattering angle, $\alpha$ and $\beta$ are constants and $r = 1,2,3,...\:$. The resummed scattering amplitude is then exactly decomposed into the sum of a nearside (N) subamplitude and a farside (F) subamplitude. We make two applications of the NF resummed theory: to elastic angular scattering in a strongly absorptive collision and to a state-to-state differential cross section for the I + HI $\rightarrow$ IH + I reaction. In both applications, we can understand the physical origin of structure in the angular scattering for suitable choices of $\alpha$, $\beta$ and $r$. 
\end{abstract}

\newpage

\section{Introduction}\label{sect:intro}

$\:\:\:\:\:$ The Fuller nearside-farside (NF) theory of elastic angular scattering has been used for more than 25 years to understand structure in the differential cross sections of nucleus-nucleus collisions. \cite{Full,FullPLB75,FullPRC75,Huss,Brink,Brand} More recently, it has been demonstrated that an extended NF theory is a powerful tool for analysing structure in the angular scattering of elastic, inelastic and reactive atom-molecule collisions. \cite{CMSS,SCS,SCS2,MC,SCS3,WMC,MCS,SC,HC,HCMol,DMC,MCS2,WNC,Nol} 

$\:\:\:\:\:$One advantage of NF theory is that the NF decomposition of the partial wave series (PWS) representation for the scattering amplitude is exact. However, this exactness is no guarantee that the NF procedure will always yield a physically meaningful explanation of structure in a differential cross section. An example is elastic scattering in a strongly absorptive collision. \cite{HC,HCMol,Hatch} It has been shown in Refs. 15 and 16 that the failure of the NF method for this type of collision can be overcome by resumming the PWS before applying the NF decomposition. The effect of the resummation is to extract from the PWS the factor $(1 - \cos \theta)^{-r}$ where $r = 1,2,3,...$ and $\theta$ is the scattering angle. \cite{HC,HCMol}

$\:\:\:\:\:$The purpose of this paper is to further develop (and apply) resummation theory for a PWS which is expanded in a basis set of Legendre polynomials. In order to motivate our development, we first consider in section II a strongly absorptive elastic collision using a simple parameterized scattering matrix element. 
We discuss an example where extracting the factor $(1 - \cos \theta)^{-r}$ results in physically unrealistic N and F cross sections. We show for this example that the correct factor to remove from the PWS is $(1 + \cos \theta)^{-r}$.
These considerations suggest that we must extend our earlier work \cite{HC,HCMol} and resum a Legendre PWS in which the more general factor $(\alpha + \beta \cos \theta)^{-r}$ is extracted, where $\alpha$ and $\beta$ are constants. This is done in section III, where we also discuss properties of the resummed PWS. We apply our resummation theory in section IV to analyse structure in the differential cross section of the state selected chemical reaction \cite{Nol,RV} 
\begin{equation}
\textrm {I + HI}(v_i=0,j_i=4) \rightarrow \textrm {IH}(v_f=0,j_f=4) \textrm { + I}  \nonumber
\end{equation}
where $v_{i},j_{i}$ and $v_{f},j_{f}$ are initial and final vibrational and rotational quantum numbers respectively.
Our conclusions are in section V.

\section{Elastic scattering in a strongly absorptive collision}

$\:\:\:\:\:$This section reports and discusses PWS, N and F differential cross sections for a strongly absorptive elastic collision. In particular, we examine how the N and F cross sections change when the factor $(\alpha + \beta \cos \theta)^{-r}$, with $r = 1,2,3,...$ is removed from the PWS for two special cases: $(\alpha, \beta) = (1,-1)$ and $(\alpha, \beta) = (1, 1)$.

\subsection{Parameterized scattering matrix element}

$\:\:\:\:\:$We use a simple two parameter analytic expression for the scattering matrix element, $S_J$, namely \cite{Hatch}

\begin{equation}\label{PSM}
S_J = \{ 1 + \exp \left[ \frac{\Lambda - \left(J + \frac{1}{2}\right) }{\Delta} \right] \} ^{-1} + \{ 1 + \exp \left[ \frac{\Lambda +\left(J + \frac{1}{2}\right)}{\Delta} \right] \} ^{-1}, \:\:\:\:\:J = 0,1,2,... 
\end{equation}

where $J$ is the total (= orbital for this case) angular momentum quantum number, $\Lambda$ locates the absorbing surface in $J$ space and $\Delta$ measures the width of the surface region.
The parameterization (\ref{PSM}) has several advantages: \cite{HC,HCMol,Hatch}

$\:\:\:\:\:$$\bullet$ For appropriate values of $\Lambda$ and $\Delta$, the angular distribution can vary by many orders of magnitude. This allows a systematic study to be made of the N and F angular scattering for many values of $r = 0,1,2,...$ .

$\:\:\:\:\:$$\bullet$ Since $S_J$ is real for $J = 0,1,2,...$, (i.e. a purely absorbing collision) the N and F cross sections are equal, which simplifies the physical interpretation of the results.

$\:\:\:\:\:$$\bullet$ The analytic expression (\ref{PSM}) is sufficiently simple that the semiclassical limit of the PWS can be worked out. \cite{Hatch} This allows the N and F components of the scattering amplitude to be unambiguously identified. \cite{Hatch}

\subsection{Examples of elastic scattering}

$\:\:\:\:\:$We start with the PWS for the elastic scattering amplitude, $f(\theta)$, written in the form

\begin{equation}\label{pws}
f(\theta) = (2ik)^{-1} \sum_{J=0}^{\infty} a_J^{(0)} P_J(\cos \theta)
\end{equation}

where $k$ is the wavenumber for the collision, $P_J(\bullet)$ is a Legendre polynomial of degree $J$, and $a_J^{(0)}$ contains information on the scattering dynamics. The significance of the superscript $^{(0)}$ will become clear from the following equations. 

$\:\:\:\:\:$It has been shown in Refs. 15, 16 and 23 that the recurrence relation obeyed by $\cos \theta P_J(\cos \theta)$ allows a resummation of eqn (\ref{pws}). We have for $\theta \neq 0$

\begin{equation}\label{rpws}
f(\theta) = (2ik)^{-1} (1 - \cos \theta)^{-r} \sum_{J=0}^{\infty} a_J^{(r)} P_J(\cos \theta),
\end{equation}
\begin{equation}
 \:\:\:\:\:\:\:\:\:\:\:\:\:\:\:\:\:\:\:\:\:\:\:\:\:\:\:\:\:\:\:\:\:\:\:\:\:\:\:\:\:\:\:\:\:\:\:\:\:\:\:\:\:\:\:\:  r = 0,1,2,..., \:\:\:\:\:\:\:\:\:\:\:\: (\alpha, \beta) = (1, -1) \nonumber
\end{equation}

where the $a_J^{(r)}$ for $r = 1,2,3,...$ are determined by the linear recurrence

\begin{equation}\label{ar}
%\begin{eqnarray}\label{ar}
a_J^{(r)} = - \frac{J}{2J - 1} a_{J-1}^{(r-1)} + a_{J}^{(r-1)} - \frac{(J + 1)}{2J + 3} a_{J+1}^{(r-1)},
\end{equation}
\begin{equation}
 \:\:\:\:\:\:\:\:\:\:\:\:\:\:\:\:\:\:\:\:\:\:\:\:\:\:\:\:\:\:\:\:\:\:\:\:\:\:\:\:\:\:\:\:\:\:\:\:\:\:\:\:\:\:\:\: J = 0,1,2,...,\:\:r = 1,2,3...,\:\:(\alpha, \beta)= (1, -1) \nonumber
%\end{eqnarray}
\end{equation}

and $a_J^{(r)} \equiv 0$ for unphysical values of $J$ i.e. $J = -1, -2, -3, ...$, when $r = 0,1,2,...\:$.

Explicit formulae for the $a_J^{(r)}$ in terms of $a_J^{(0)}$ when $r = 1,2,\: \rm and \: 3$ have been given in Ref. 16;
they can also be obtained from our results in section III.B as a special case.

$\:\:\:\:\:$The NF decomposition for the resummed scattering amplitude (\ref{rpws}) is obtained by writing \cite{FullPRC75}

\begin{equation}\label{P=Qpm}
P_J(\cos \theta) = Q_J^{(+)} (\cos \theta) + Q_J^{(-)} (\cos \theta)
\end{equation}

where (for $\theta \neq 0, \pi$)

\begin{equation}\label{Qpm}
Q_J^{(\pm)} (\cos \theta) = \frac{1}{2} \left[ P_J(\cos \theta) \mp (2i/ \pi) Q_J(\cos \theta) \right]
\end{equation}

In eqn (\ref{Qpm}), $Q_J(\bullet)$ is a Legendre function of the second kind. Substituting eqn (\ref{P=Qpm}) into eqn (\ref{rpws}) gives
\begin{equation}\label{f=pm}
f(\theta) = f_r^{(+)} (\theta) + f_r^{(-)} (\theta), \:\:\: r = 0,1,2,...
\end{equation}

where

\begin{equation}\label{fpm}
f_r^{(\pm)} (\theta) = (2ik)^{-1} (1 - \cos \theta)^{-r} \sum_{J=0}^{\infty} a_J^{(r)} Q_J^{(\pm)}(\cos \theta),
\end{equation}
\begin{equation}
 \:\:\:\:\:\:\:\:\:\:\:\:\:\:\:\:\:\:\:\:\:\:\:\:\:\:\:\:\:\:\:\:\:\:\:\:\:\:\:\:\:\:\:\:\:\:\:\:\:\:\:\:\:\:\:\:  r = 0,1,2,..., \:\:\:\:\:\:\:\:\:\:\:\: (\alpha, \beta) = (1, -1) \nonumber
\end{equation}

$\:\:\:\:\:$The asymptotic forms of $P_J(\cos \theta),\: Q_J(\cos \theta)$ and $Q_J^{(\pm)}(\cos \theta)$ for $J \sin \theta \gg 1$ show that $f_r^{(-)}(\theta)$ is a N subamplitude and $f_r^{(+)}(\theta)$ is a F subamplitude. \cite{FullPRC75,MC,HCMol} The corresponding PWS, N and F differential cross sections are given by
\begin{equation}\label{I=fsq}
I(\theta) = |f(\theta)|^{2}
\end{equation}

and

\begin{equation}\label{I=fsqpm}
I_r^{(\pm)}(\theta) = |f_r^{(\pm)}(\theta)|^{2}, \:\:\:r = 0,1,2,...
\end{equation}

Note that the N and F subamplitudes (\ref{fpm}) depend on $r$, whereas their sum (\ref{f=pm}), $f(\theta)$, does not.

$\:\:\:\:\:$Our first choice for $a_J^{(0)}$ in the PWS (\ref{pws}) is \cite{HC,HCMol,Hatch}

\begin{equation}\label{a=2j+1S}
a_J^{(0)} = (2J + 1) (S_J - 1), \:\:\: J = 0,1,2,...
\end{equation}

Figure 1 shows plots of the dimensionless quantities $\rm ln \left[ \it k^{\rm 2} \it I(\theta) \right]$ and $\rm ln \left[ \it k^{\rm 2} \it I_r^{(+)}(\theta) \right] \equiv \rm ln \left[ \it k^{\rm 2} \it I_r^{(-)}(\theta) \right]$, with $r = 0,1,2, \: \rm and \: 3$ for $\Lambda = 40$, $\Delta = 5$ in eqns (\ref{PSM}) - (\ref{a=2j+1S}). A precision of about 34 significant digits is necessary to generate the results in Figure 1 using a maximum cut-off value of $J_{max} = 500$. It can be seen that the PWS cross section possesses oscillations over the whole angular range. Semiclassically, these oscillations arise from interference between the N and F subamplitudes. \cite{Hatch} The corresponding semiclassical N and F cross sections are oscillation free (not illustrated in Fig. 1, but displayed in Fig. 2a of Ref. 21). Figure 1 shows that the angular range over which the NF subamplitudes (\ref{fpm}) provide a physically meaningful interpretation of the oscillations gets larger as $r$ increases. \cite{HCMol} The NF cross sections also converge toward the semiclassical NF cross sections as $r$ increases. The values $(\alpha, \beta)= (1,-1)$ are therefore a satisfactory choice in the NF resummed theory. Some additional discussion of the cross sections in Fig. 1 can be found in Refs. 15, 16 and 21. 

$\:\:\:\:\:$Our second choice for $a_J^{(0)}$ is 

\begin{equation}\label{a=2j+1-1S}
a_J^{(0)} = (2J + 1) (-1)^J (S_J - 1),  \:\:\:\:\:\:  J = 0,1,2,...
\end{equation}

and we again use $\Lambda = 40$, $\Delta = 5$ and $(\alpha, \beta)= (1,-1)$ in eqns (\ref{PSM}) - (\ref{I=fsqpm}), (\ref{a=2j+1-1S}). Figure 2 shows that the PWS angular distribution is now backward peaked. Unfortunately, the NF cross sections increasingly diverge for $\theta \lesssim 175 ^\circ$ as $r$ changes from $r = 0$ to $r = 1,2,3$, i.e. the NF decomposition no longer provides a physically meaningful interpretation of the oscillations in the PWS angular distribution for $\theta \lesssim 175 ^\circ$. 

$\:\:\:\:\:$The PWS cross section plotted in Fig. 2 is the reflection across $ \theta = \pi/2$ of the PWS curve in Fig. 1. This property is a consequence of the identity $(-1)^J P_J(\cos \theta) = P_J(\cos (\pi - \theta))$. 
The same reflection property is also true for the $r = 0$ NF cross sections because $(-1)^J Q_J(\cos \theta) = - Q_J(\cos (\pi - \theta))$. 
These observations suggest that the correct factor to remove from the PWS for the NF decomposition to be physically successfull is $(1 - \cos(\pi - \theta))^{-r}$ i.e. $(1 + \cos \theta)^{-r}$. We can again use the recurrence obeyed by $\cos \theta P_J(\cos \theta)$  
to resum eqn (\ref{pws}) to remove the factor $(1 + \cos \theta)^{-r}$ for $\theta \neq \pi$. We find that eqns (\ref{rpws}), (\ref{ar}) and (\ref{fpm}) are replaced by the following results
\begin{eqnarray}\label{fpwsab}
f(\theta) = (2ik)^{-1} (1 + \cos \theta)^{-r} \sum_{J=0}^{\infty} a_J^{(r)} P_J(\cos \theta), \\ \:\:\:\:\:\:\:\:\:r = 0,1,2,... \:,\:\:(\alpha, \beta) = (1, 1) \nonumber
\end{eqnarray}

where the $a_J^{(r)}$ obey the linear recurrence
\begin{eqnarray}\label{acoeff}
a_J^{(r)} = \frac {J}{2J - 1} a_{J-1}^{(r-1)} + a_J^{(r-1)} + \frac {(J + 1)}{2J + 3} a_{J+1}^{(r-1)}, \:\:\:\:\:\:\:\:\:\:\:\:\:\:\:\:\:\:\:\:\:\:\:\:\: \\ \:\:\:\:\:\:\:\:\:\:\:\:\:\:\:\:\:\:\:\:\:\:\:\:\:\:\:\:\:\:\:\:\:\:\:\:\:\:\:\:\:\:\:\:\:\:\:\:\:\:\:\:\:\:\:\:\:\:J = 0,1,2,...\:, \:\:\:r = 1,2,3,...\:, \: (\alpha, \beta) = (1,1) \nonumber
\end{eqnarray}

and $a_J^{(r)} \equiv 0$, for unphysical values of $J$ when $r = 0,1,2,...\:$. The NF subamplitudes are given by

\begin{eqnarray}\label{pws2}
f_r^{(\pm)}(\theta) = (2ik)^{-1} (1 + \cos \theta)^{-r} \sum_{J=0}^{\infty} a_J^{(r)} Q_J^{(\pm)}(\cos \theta), \\ \:\:\:\:\:\:\: r=0,1,2,..., \:\:\:(\alpha, \beta) = (1,1) \nonumber
\end{eqnarray}

Figure 3 shows the PWS and NF angular distributions when $\Lambda = 40$ and $\Delta = 5$ are used in eqns (\ref{a=2j+1-1S}) - (\ref{pws2}) for $r = 0,1,2, \: \rm and \: 3$ . It can be seen that, in contrast to Fig. 2, the NF cross sections now provide a physically meaningful interpretation of the oscillations over an increasingly wider angular range as $r$ increases. In fact, Fig. 3 is just the reflection of Fig. 1 across $\theta = \pi/2$.

$\:\:\:\:\:$The simple examples discussed in this section show that we must extend the resummation theory developed in Refs. 15 and 16 in order to remove the general factor $(\alpha + \beta \cos \theta)^{-r}$ from the PWS.

\section{Resummation of Partial Wave Series}

$\:\:\:\:\:$In this section, we show how to resum a Legendre PWS so as to extract the general factor $(\alpha + \beta \cos \theta)^{-r}$ with $r = 1,2,3,...$. In the following manipulations, $\alpha$ and $\beta$ can be complex numbers, although in all our applications $\alpha$ and $\beta$ are real. We also derive explicit formulae for the coefficients of the resummed series $a_J^{(r)} (\alpha,\beta)$ in terms of $a_J^{(0)}$ and $S_J$ for $r = 1,2,3$ and discuss some of their properties.

\subsection{Resummation of the scattering amplitude}

We start with the PWS for $f(\theta)$ written in the more compact form
\begin{equation}\label{31}
2ikf(\theta)=\sum_{J=0}^{\infty} a_J^{(0)}
P_J(x),\:\:\:\:\:\:\:\:x=\cos \theta 
\end{equation}
Multiplication of eqn (\ref{31}) by ${\alpha + \beta x \neq 0}$ gives
\begin{equation}\label{32}
2ik(\alpha + \beta x)f(\theta)= \alpha \sum_{J=0}^{\infty} a_J^{(0)} P_J(x) + \beta \sum_{J=0}^{\infty} a_J^{(0)} x P_J(x)
\end{equation}
Next we apply the recurrence relation
\begin{equation}\label{33}
(2J+1)xP_J(x)= JP_{J-1}(x) + (J+1)P_{J+1}(x),\:\:\:\:\:\:\:\:J=0,1,2,...,
\end{equation}
to the second term on the r.h.s. of eqn (\ref{32}) obtaining
\begin{equation}\label{34}
\sum_{J=0}^{\infty}a_J^{(0)}x P_J(x)= \sum_{J=0}^{\infty} a_J^{(0)}\frac{J}{2J+1} P_{J-1}(x) + \sum_{J=0}^{\infty} a_J^{(0)}\frac{(J+1)}{2J+1} P_{J+1}(x). 
\end{equation}
An important point for the following derivation is that the recurrence (\ref{33}) is valid for $J = 0$ as well as for $J = 1,2,3,...$.

$\:\:\:\:\:$We can manipulate the first series on the r.h.s. of eqn (\ref{34}) as follows:

\begin{eqnarray}
\sum_{J=0}^{\infty} a_J^{(0)} \frac{J}{2J+1} P_{J-1}(x) \:\:\:\:\:\:\:\:\:\:\:\:\:\:\:\:\:\:\:\:\:\:\:\:\:\:\:\:\:\:\:\:\:\:\:\:\:\:\:\:\:\:\:\:\:\:\:\:\:\:\:\:\:\:\:\:\:\:\:\:\:\:\:\:\:\:\:\:\:\:\:\:\:\:\:\:\:\:\:\:\:\:\:\:\:\:\:\:\:\:\:\:\:\:\:\:\:\:\:\:\:\:\:\:\:\:\:\:\:\:\:\:\:\:\:\:\:\:\:\:\:\:\:\:\:\:\:\:\:\:\:\: \nonumber \\
\:\:\:=\sum_{J=1}^{\infty} a_J^{(0)} \frac{J}{2J+1} P_{J-1}(x) \:\:\:\text{since}\:\:\:\:JP_{J-1}(x)=0\:\:\:\:\text{for}\:\:\:\:J=0, \:\:\:\:\:\:\:\:\:\:\:\:\:\:\:\:\:\:\:\:\:\:\:\:\:\:\:\:\:\: \nonumber \\
\:\:\:\:\:\:\:\:\:=\sum_{J=0}^{\infty} a_{J+1}^{(0)} \frac{(J+1)}{2J+3}P_{J}(x)\:\:\:\:\text{after replacing}\:\:\:\:J-1\:\:\:\:\text{by}\:\:\:\:J^\prime \:\:\:\:\text{and}\:\:\:\: J^\prime \rightarrow J. \:\:\:\:\: \nonumber
\end{eqnarray}

Similarly for the second series on the r.h.s. of eqn (\ref{34}) we have

\begin{eqnarray}
\sum_{J=0}^{\infty} a_J^{(0)} \frac{(J+1)}{2J+1} P_{J+1}(x) \:\:\:\:\:\:\:\:\:\:\:\:\:\:\:\:\:\:\:\:\:\:\:\:\:\:\:\:\:\:\:\:\:\:\:\:\:\:\:\:\:\:\:\:\:\:\:\:\:\:\:\:\:\:\:\:\:\:\:\:\:\:\:\:\:\:\:\:\:\:\:\:\:\:\:\:\:\:\:\:\:\:\:\:\:\:\:\:\:\:\:\:\:\:\:\:\:\:\:\:\:\:\:\:\:\:\:\:\:\:\:\:\:\:\:\:\:\:\:\:\:\:\:\:\:\:\:\:\:\:\:\: \nonumber \\
\:\:\:=\sum_{J=-1}^{\infty} a_J^{(0)} \frac{(J+1)}{2J+1} P_{J+1}(x) \:\:\:\:\text{since}\:\:\:\:(J+1)P_{J+1}(x)=0\:\:\:\:\text{for}\:\:\:\:J=-1, \:\:\:\:\:\:\:\:\:\:\: \nonumber \\
\:\:\:\:\:\:\:\:\:=\sum_{J=0}^{\infty} a_{J-1}^{(0)} \frac{J}{2J-1}P_{J}(x)\:\:\:\:\text{after replacing}\:\:\:\:J+1\:\:\:\:\text{by}\:\:\:\:J^\prime  \:\:\:\:\text{and}\:\:\:\: J^\prime \rightarrow J. \:\:\:\:\:\: \nonumber
\end{eqnarray}

Combining the above results lets us write eqn (\ref{32}) in the form
\begin{equation}\label{35}
2ik(\alpha + \beta x)f(\theta)=\sum_{J=0}^{\infty} a_J^{(1)} (\alpha,\beta) \: P_J(x) 
\end{equation}
where
\begin{equation}\label{36}
a_J^{(1)}(\alpha,\beta) =\beta \frac{J}{2J-1}a_{J-1}^{(0)} +\alpha a_{J}^{(0)}
+ \beta \frac{(J+1)}{2J+3}a_{J+1}^{(0)} \:, \:\:\:\:\:\:\:\:J=0,1,2,..., 
\end{equation}
We can again multiply eqn (\ref{35}) by $\alpha + \beta x$ and repeat the above
procedure. The general result for $\alpha + \beta x \neq 0$ is
\begin{eqnarray}\label{37}
2ikf(\theta)=(\alpha + \beta x)^{-r} \sum_{J=0}^{\infty} a_J^{(r)} (\alpha,\beta) P_J(x),
\:\:\:\:\:\:\:\:r=0,1,2,...,
\end{eqnarray}
where the $a_J^{(r)}(\alpha,\beta)$ satisfy the linear recurrence
\begin{eqnarray}\label{38}
a_J^{(r)} (\alpha,\beta)=\beta \frac{J}{2J-1}a_{J-1}^{(r-1)}(\alpha,\beta) + \alpha a_{J}^{(r-1)} (\alpha,\beta) + \beta \frac{(J+1)}{2J+3}a_{J+1}^{(r-1)} (\alpha,\beta), \\
\:\:\:\:\:\:\:\:\:\:\:\:\:\:\:\:\:\:\:\:\: r=1,2,3,...,  J=0,1,2,..., \nonumber
\end{eqnarray}
with $a_J^{(0)} \equiv a_J^{(0)}(\alpha,\beta)$ and $a_J^{(r)}(\alpha,\beta) \equiv 0$ for $J=-1,-2,-3,...\:$ when $r = 0,1,2,... \:$.
Special cases of eqn (\ref{38}) are eqn (\ref{ar}) when $(\alpha,\beta) = (1,-1)$ and eqn (\ref{acoeff}) when $(\alpha,\beta) = (1,1)$. Explicit formulae for the $a_J^{(r)}(\alpha,\beta)$ with $r = 1,2, \: \rm and \: 3$ in terms of $a_J^{(0)}$ and $S_J$ are given in section III.B.

$\:\:\:\:\:$The NF decomposition of the resummed scattering amplitude is
\begin{equation}\label{fmf}
f(\theta)=f_r^{(+)}(\alpha, \beta ; \theta)+f_r^{(-)}(\alpha, \beta ; \theta),\:\:\:\:r=0,1,2,...,
\end{equation}
where the NF resummed subamplitudes are
\begin{equation}\label{fmfpws}
f_r^{(\pm)}(\alpha, \beta ; \theta)=(2ik)^{-1} (\alpha + \beta \cos \theta)^{-r}\sum_{J=0}^\infty a_J^{(r)}(\alpha,\beta) Q_J^{(\pm)}(\cos \theta),\:\:\:\:r=0,1,2,...,
\end{equation}

The corresponding NF resummed differential cross sections are given by
\begin{equation}\label{311}
I_r^{(\pm)}(\alpha, \beta ; \theta)= | f_r^{(\pm)}(\alpha, \beta ; \theta) |^2, \:\:\:\:r=0,1,2,...,
\end{equation}

When $r$ = 0, there is no dependance on $\alpha$ and $\beta$ in eqns (\ref{fmf}) - (\ref{311}).
\subsection{Explicit formulae for ${a_J^{(r)}}(\alpha,\beta)$ when $r$ = 1,2, and 3.}
This section lists explicit formulae for the $a_J^{(r)}(\alpha,\beta)$ in terms of (a)
$a_J^{(0)}$ and (b) $S_J$ for $r = 1,2,\: \rm and \:3$. We used the algebraic software
package $Mathematica$ 3.0 to generate \cite{Wolf} the required
formulae from the defining eqns (\ref{36}) and (\ref{38}).

(a) The formulae expressing $a_J^{(r)}(\alpha,\beta)$ in terms of $a_J^{(0)}$, where
$J=0,1,2,...,$ are

\begin{eqnarray}
a_J^{(1)}(\alpha,\beta) &=& \beta \frac{J}{2J-1} a_{J-1}^{(0)} + \alpha a_J^{(0)}
+ \beta \frac{(J+1)}{2J+3}a_{J+1}^{(0)} \nonumber \\
\nonumber \\
a_J^{(2)}(\alpha,\beta) &=& \beta^2 \frac{J(J-1)}{(2J-1)(2J-3)} a_{J-2}^{(0)} +2 \alpha \beta \frac{J}{2J-1} a_{J-1}^{(0
)} \nonumber \\ 
&\:&+ \left[ \alpha^2 + \beta^2 \frac{(2J^2 + 2J -1)}{(2J+3)(2J-1)}\right] a_{J}^{(0)} \nonumber \\
&\:&+ 2\alpha\beta \frac{(J+1)}{2J+3} a_{J+1}^{(0)} + \beta^2 \frac{(J+2)(J+1)}{(2J+5)(2J+3)} a_{J+2}^{(0)} \nonumber \\
\nonumber \\
a_J^{(3)}(\alpha,\beta) &=& \beta^3 \frac{J(J-1)(J-2)}{(2J-1)(2J-3)(2J-5)} a_{J-3}^{(0)} + 3 \alpha \beta^2 \frac{J(J-1)}{(2J-1)(2J-3)} a_{J-2}^{(0)} \nonumber \\
&\:&+ 3 \left[ \alpha^2 \beta \frac{J}{2J-1} + \beta^3 \frac {J(J^2-2)}{(2J+3)(2J-1)(2J-3)} \right] a_{J-1}^{(0)} \nonumber \\
&\:&+ \left[ \alpha^3 + 3 \alpha \beta^2 \frac {(2J^2+2J-1)}{(2J+3)(2J-1)} \right] a_{J}^{(0)} \nonumber \\ 
&\:&+ 3 \left[ \alpha^2 \beta \frac {(J+1)}{2J+3} + \beta^3\frac {(J+1)(J^2+2J-1)}{(2J+5)(2J+3)(2J-1)} \right] a_{J+1}^{(0)} \nonumber \\
&\:&+3 \alpha \beta^2 \frac{(J+2)(J+1)}{(2J+5)(2J+3)} a_{J+2}^{(0)} + \beta^3 \frac {(J+3)(J+2)(J+1)}{(2J+7)(2J+5)(2J+3)}a_{J+3}^{(0)} \nonumber 
\end{eqnarray}

$\:\:\:\:\:$The above formulae appear to require values for the non-physical coefficients $a_{-1}^{(0)}$, $a_{-2
}^{(0)}$, $a_{-3}^{(0)}$
e.g. when $J=0$. However these coefficients are always multiplied by
terms that are zero, so they do not contribute, i.e. we can always set
$a_J^{(r)}(\alpha,\beta) \equiv 0$ when $J<0$ for $r = 0,1,2,...\:$.
As an example, the resummed coefficients for $J = 0$ are given by
\begin{eqnarray}
a_0^{(1)}(\alpha,\beta) &=& \alpha \: a_0^{(0)} + \frac {1}{3} \:\beta \:a_1^{(0)} \nonumber \\
a_0^{(2)}(\alpha,\beta) &=& (\alpha^2 + \frac {1}{3} \: \beta^2) a_0^{(0)} + \frac {2}{3} \: \alpha \: \beta a_1^{(0)} + \frac {2}{15} \: \beta^2 a_2^{(0)} \nonumber \\
a_0^{(3)}(\alpha,\beta) &=& (\alpha^3 + \alpha \beta^2) a_0^{(0)} + (\alpha^2 \beta + \frac {1}{5} \: \beta^3) a_1^{(0)} + \frac {2}{5} \: \alpha \: \beta^2 a_2^{(0)} + \frac {2}{35} \: \beta^3 a_3^{(0)} \nonumber \\
\end{eqnarray}
(b) In order to handle both elastic and inelastic (or reactive) scattering, we write
\begin{equation}
a_J^{(0)} = (2J + 1)(S_J - \delta)
\end{equation}

where for elastic scattering, the delta function, $\delta = 1$ and $S_J \rightarrow 1$ as $J \rightarrow \infty$, whereas for inelastic scattering, $\delta = 0$ and $S_J \rightarrow 0$ as $J \rightarrow \infty$.
The formulae for $a_J^{(r)}$ expressed in terms of
$S_J$, where $J=0,1,2,...,$ are:

\begin{eqnarray}
a_J^{(1)}(\alpha,\beta) &=& \beta \: J \: S_{J-1}+\alpha \: (2J+1) \: S_J+\beta \: (J+1) \: S_{J+1} - (\alpha + \beta)(2J+1)\delta \nonumber \\
\nonumber \\
a_J^{(2)}(\alpha,\beta) &=&
\beta^2 \frac{J(J-1)}{2J-1} S_{J-2} + 2 \alpha\beta J S_{J-1} \nonumber \\
&\:&+ (2J+1) \left[ \alpha^2 + \beta^2 \frac{(2J^2+2J-1)}{(2J+3)(2J-1)} \right] S_{J} \nonumber \\
&\:&+ 2 \alpha\beta(J+1)S_{J+1} + \beta^2 \frac{(J+2)(J+1)}{2J+3} S_{J+2} - (\alpha + \beta)^2 (2J+1)\delta
\nonumber \\
\nonumber \\
a_J^{(3)}(\alpha,\beta) &=& \beta^3\frac{J(J-1)(J-2)}{(2J-1)(2J-3)}S_{J-3}
+3\alpha\beta^2\frac{J(J-1)}{2J-1}S_{J-2} \nonumber \\
&\:&+ 3J \left[ \alpha^2\beta + \beta^3\frac{(J^2-2)}{(2J+3)(2J-3)} \right] S_{J-1} \nonumber \\
&\:&+ (2J+1) \left[ \alpha^3 + 3 \alpha\beta^2\frac{(2J^2+2J-1)}{(2J+3)(2J-1)} \right] S_J \nonumber \\
&\:&+ 3(J+1) \left[ \alpha^2\beta + \beta^3\frac{(J^2+2J-1)}{(2J+5)(2J-1)} \right] S_{J+1} \nonumber \\
&\:&+ 3\alpha\beta^2\frac{(J+2)(J+1)}{2J+3}S_{J+2} + \beta^3\frac{(J+3)(J+2)(J+1)}{(2J+5)(2J+3)}S_{J+3} \nonumber \\
&\:& - (\alpha + \beta)^3 (2J+1) \delta \nonumber
\end{eqnarray}
By the same reasoning as before, we can set $S_J\equiv0$ for $J<0$. For elastic scattering, where $\delta = 1$, notice that the terms involving the delta function only vanish if $\alpha = - \beta$. For $J=0$, the above equations simplify to  
\begin{eqnarray}
a_0^{(1)}(\alpha,\beta) &=& \alpha \: S_0 + \beta \: S_1 - (\alpha + \beta) \: \delta \nonumber \\
a_0^{(2)}(\alpha,\beta) &=& (\alpha^2 + \frac {1}{3} \: \beta^2) \: S_0 + 2 \: \alpha \: \beta \: S_1 + \frac {2}{3} \: \beta^2 \: S_2 - (\alpha + \beta)^2 \: \delta \nonumber \\
a_0^{(3)}(\alpha,\beta) &=& (\alpha^3 + \alpha \: \beta^2)\: S_0 + 3 \: (\alpha^2 \: \beta + \frac {1}{5} \: \beta^3) \: S_1 + 2\: \alpha \: \beta^2 \: S_2 + \frac {2}{5} \: \beta^3 \: S_3 - (\alpha + \beta)^3 \: \delta \nonumber
\end{eqnarray}

\subsection{Discussion}

$\:\:\:\:\:$We make the following remarks on the results derived in section III.B for $r = 1,2,3,...$ (always assuming that $\alpha + \beta \cos \theta \neq 0$):

$\:\:\:\:\:$ $\bullet$ $\alpha \neq 0,\: \beta = 0$. This case is trivial in that eqn (\ref{37}) for $r = 1,2,3,...$ immediately reduces to eqn (\ref{31}) for $r = 0$.

$\:\:\:\:\:$ $\bullet$ $\alpha = 0,\: \beta \neq 0$. This case has the possible disadvantage for numerical work that the factor $(\beta \cos \theta)^{-r}$ becomes singular as $\theta \rightarrow \pi/2$.

$\:\:\:\:\:$ $\bullet$ $\alpha \neq 0,\: \beta \neq 0$. For a given value of $r$, the denominator, $(\alpha + \beta \cos \theta)^r$, and the resummed coefficient, $a_J^{(r)} (\alpha,\beta)$, are both homogeneous functions of $\alpha$ and $\beta$ of degree $r$. This implies the resummation theory can be developed in terms of the single parameter $\gamma = \alpha/\beta$.

$\:\:\:\:\:$ $\bullet$ The numerical value of $f(\theta)$, as given by eqn (\ref{37}), is of course independent of the values chosen for $\alpha,\beta$ and $r$. This allows a valuable check that the resummed coefficients have been correctly programmed on a computer.

$\:\:\:\:\:$ $\bullet$ In our applications in sections II and IV, $\alpha$ and $\beta$ are always chosen to be real (see also Ref. 19). In addition, we require the condition $\alpha + \beta \cos \theta \neq 0$ to hold not just at a single angle, but for all $\theta \in (0, \pi)$. This implies the restriction $|\alpha| \geqslant |\beta|$.

$\:\:\:\:\:$ $\bullet$ For $(\alpha,\beta) = (1,-1)$, Wimp \cite{Wimp} has used Wilf-Zeilberg algorithms to study the mathematical properties of the $a_J^{(r)}(1,-1)$. He concluded that no simple closed form exists for them, i.e. their evaluation requires the explicit calculations described in section III.B and in Ref. 16. 

\section{Angular scattering for the I + HI $\rightarrow$ IH + I reaction}
$\:\:\:\:\:$ We have previously used \cite{Nol} the unresummed $r = 0$ NF theory to analyse angular scattering for the state-to-state reaction
\begin{equation}
\rm {I + HI}(\it {v_i} = \rm 0, \it {j_i} \leqslant \rm 5) \rightarrow \rm {IH} (\it {v_f} = \rm 0, \it {j_f} \leqslant \rm 5) + \rm I \nonumber
\end{equation}
on the extended London-Eyring-Polanyi-Sato potential energy surface A of Manz and R\"omelt. \cite{Manz}

$\:\:\:\:\:$We demonstrated that the $r = 0$ NF decomposition nearly always provides a physically clear explanation of the forward, sideward and backward scattering. \cite{Nol} However, in a few cases, the physical interpretation was obscured by the presence of oscillations in the N and F cross sections at forward angles. \cite{Nol}

$\:\:\:\:\:$An example is shown in Fig. 4 which displays PWS, N and F angular distributions for the $j_i = j_f = 4$ transition at a total energy of $E$ = 29.5 meV, where $E$ = 0 meV corresponds to the energy of HI($v_i = 0, j_i =0$). The scattering matrix elements calculated in Ref. 20 were used to generate Fig. 4. They were computed by a quantum method which applies a Born-Oppenheimer type separation to the motion of the light and heavy atoms (a centrifugal sudden approximation is also made). \cite{RV}

$\:\:\:\:\:$ Note that Fig. 4 uses the reactive scattering angle $\theta_R$ along the abscissa, which is defined as the angle between the direction of the outgoing IH molecule and the incoming I atom. It is related to the angle $\theta$ employed in section III by $\theta_R = \pi - \theta$.

$\:\:\:\:\:$The PWS for the reactive scattering amplitude is given by \cite{Nol,RV}
\begin{equation}\label{4pws}
f(\theta_R) = (2ik)^{-1} \sum_{J=0}^{\infty} (2J + 1) \tilde{S}_J P_J(\cos \theta_R)
\end{equation} 

where $\tilde{S}_J = (-1)^J \: S_J$. [ n.b. in Refs. 20 and 22, $\tilde{S}_J$ is denoted $S_J$ and in Ref. 22, $\theta_R$ is denoted $\theta$ ]. The corresponding differential cross section is
\begin{equation}\label{4I=f2}
I(\theta_R)=|f(\theta_R)|^2
\end{equation} 

In eqns (\ref{4pws}) and (\ref{4I=f2}), we have omitted the subscript $\it v_fj_f \leftarrow v_ij_i$ from $\tilde{S}_J$, $f(\theta_R)$ and $I(\theta_R)$, as well as the subscript $\it v_ij_i$ from $k$ = 20.2$\: \rm \AA^{-1}$ in order to keep the notation simple, since we always have $\it v_i$ = 0, $\it j_i$ = 4 and $\it v_f$ = 0, $\it j_f$ = 4 in our calculations. The masses used are 1.008 u for the H atom and 126.9 u for the I atom. N and F resummed subamplitudes and cross sections can also be defined, which are the same as eqns (\ref{fmf}) - (\ref{311}) provided the changes $S_J \rightarrow \tilde{S}_J$ and $\theta \rightarrow \theta_R$ are made.

$\:\:\:\:\:$ Figure 4 shows that the PWS angular distribution is N dominated for $\theta_R \gtrsim 60 ^\circ$. At smaller angles, there are high frequency diffraction oscillations of period $\Delta \theta_R \approx 2.6 ^\circ$ which arise from NF interference. However, the N and F cross sections themselves possess oscillations of period $\Delta \theta_R \approx 5.2 ^\circ$. A natural question to ask is: Are these oscillations artefacts of the $r$ = 0 NF decomposition or are they physically meaningful?

$\:\:\:\:\:$To begin to answer this question, we show in Fig. 5 the $r$ = 1,2, and 3, NF and PWS angular distributions using $(\alpha,\beta) = (1,-1)$ in the reactive analogs of eqns (\ref{fmf}) - (\ref{311}). The large angle scattering stays N dominated, but in the forward angle region, the N and F cross sections rapidly grow in magnitude; this growth becomes more pronounced as $r$ increases from $r$ = 1 to $r$ = 3. Unfortunately this behaviour is meaningless as a physical explanation of the diffraction oscillations (even though the NF decompositions are mathematically exact). The blow up at small angles can be understood from the identity \cite{HCMol}
\begin{equation}\label{4fpmab}
f_r^{(\pm)}(1,-1; \theta_R) = f_0^{(\pm)}(1,-1; \theta_R) \mp \frac{1}{2 \pi k} \sum_{s = 0}^{r-1} \frac {a_0^{(s)}(1, -1)}{(1 - \cos \theta_R)^{s + 1}} \:\:,  
\end{equation}
\begin{equation}
\:\:\:\:\:\:\:\:\:\:\:\:\:\:\:\:\:\:\:\:\:\:\:\:\:\:\:\:\:\:\:\:\:\:\:\:\:\:\:\:\:\:\:\:\:\:\:\:\:\:\:\:\:\:\:\:\:\:\:\:\:\:\:\:\:\:\:\:\:\:\:\:\:\:\: r = 0,1,2,... \nonumber
\end{equation}

where the sum is interpreted as 0 for $r$ = 0. Equation (\ref{4fpmab}) shows that the term 1 / $(1 - \cos \theta_R)^{s + 1}$ for $s$ = 0,1,2,..., $r -1$ will give rise to increasingly divergent NF cross sections at small $\theta_R$ as $r$ increases, provided the $a_0^{(s)}(1,-1)$ are not extremely small in magnitude. For the present example, the values are 

\begin{eqnarray}
a_0^{(0)}  &=& 0.0561\:  - 0.323 \it \: i  \nonumber \\
a_0^{(1)} (1, -1) &=& 0.103 \:  - 0.648 \it \: i \nonumber \\
a_0^{(2)} (1, -1) &=& 0.186 \:  - 1.299 \it \: i \nonumber \\
a_0^{(3)} (1, -1) &=& 0.334 \:  - 2.601 \it \: i \nonumber
\end{eqnarray}

$\:\:\:\:\:$Figure 6 and 7 show plots of Re $a_J^{(r)}(1,-1)$ versus $J$ and Im $a_J^{(r)}(1,-1)$ versus $J$ respectively for $r$ = 0,1,2 and 3. There is no apparent improvement in the convergence of the partial wave series for the NF resummed subamplitudes; in fact the magnitudes of the $a_J^{(r)}(1,-1)$ become larger as $r$ increases.

$\:\:\:\:\:$In summary, we can say that the NF resummed theory using $(\alpha,\beta) = (1,-1)$ has not improved the physical interpretation of the forward angle PWS diffraction oscillations -  indeed it has made matters worse.

$\:\:\:\:\:$ Figure 8 shows plots for $r$ = 1,2 and 3 of N, F and PWS differential cross sections using $(\alpha,\beta) = (1.05,1)$ in the reactive analogs of eqns (\ref{fmf}) - (\ref{311}). The N and F cross sections are now oscillation free at forward angles and provide a clearer physical interpretation of the diffraction oscillations as a NF interference effect than does the $r$ = 0 NF analysis. Figures 9 and 10 plot the real and imaginary parts respectively of $a_J^{(r)}(1.05,1)$ versus $J$ for $r$ = 0,1,2 and 3. It can be seen that the magnitudes of the $a_J^{(r)}$(1.05, 1) become smaller at low $J$ as $r$ increases. For example, at $J$ = 0 we have 
\begin{eqnarray}
a_0^{(0)} &=& 0.0561 \: - 0.323 \it \: i \nonumber \\
a_0^{(1)} (1.05, 1) &=& 0.0123 \: - 0.0147 \it \: i \nonumber \\
a_0^{(2)} (1.05, 1) &=& 0.00103 \: - 0.000103 \it \: i \nonumber \\
a_0^{(3)} (1.05, 1) &=& 0.0000190 \: - 0.0000504 \it \: i \nonumber
\end{eqnarray}

$\:\:\:\:\:$Figure 9 and 10 show that, in effect, numerically significant terms have been moved away from low values of $J$ to larger values of $J$. As discussed in Refs. 15, 16 and 19, this concentrating effect, which emphasizes partial waves with $J \gg 1$ as $r$ increases, favours a physically meaningful NF analysis because the $Q_J^{(\pm)}(\cos \theta_R)$ become travelling angular waves in this limit. In particular, we have for $J \sin \theta_R \gg 1$ 
\begin{equation}\label{4Q}
Q_J^{(\pm)}(\cos \theta_R) \sim [ 2 \pi (J + \frac{1}{2}) \sin \theta_R ]^{-1/2} \exp \{ \pm i [(J + \frac{1}{2}) \theta_R - \frac{1}{4} \pi] \} \nonumber
\end{equation}

$\:\:\:\:\:$Note that $(1.05\: +\: \cos \theta_R)^{-r}$ approximately mimics the shape of $\it I(\theta_R)$ in that both are backward peaked. This observation, together with the results discussed in section II, are examples of a rule of thumb \cite{WNC} for choosing ($\alpha,\beta$) so as to obtain physically meaningful N and F subamplitudes.

$\:\:\:\:\:$ We have also calculated N and F resummed cross sections for some other values of $(\alpha,\beta)$. For $(\alpha,\beta) = (1,1)$, we obtained similar results to Figs. 8 - 10. One difference is a blowing up of the N and F angular distributions at backward angles as $r$ increases, which is similar to, although less pronounced, than the effect in the forward direction of Fig. 5. For $(\alpha,\beta)$ = (1.5, 1), our results are similar to those for $(\alpha,\beta)$ = (1.05, 1).
\section{Conclusions}
$\:\:\:\:\:$We have shown how to remove the factor $(\alpha + \beta \cos \theta)^{-r}$ with $r$ = 1,2,3... from a Legendre PWS. We then decomposed the resummed PWS for the scattering amplitude into N and F subamplitudes. Two applications of this NF resummed theory were reported: to elastic angular scattering in a strongly absorptive collision and to a state-to-state differential cross section for the I + HI $\rightarrow$ IH + I reaction. In both applications, by making suitable choices for $(\alpha,\beta)$ and $r$, we were able to explain structure in the angular distributions. 

\section*{Acknowledgements}
$\:\:\:\:\:$We thank C. Kubach (Paris) and N. Rougeau (Paris) for providing us with scattering matrix elements for the I + HI reaction. This research has been supported by the Engineering and Physical Sciences Research Council (UK) and by INTAS (EU).

\clearpage

\begin{figure}[p]
  \begin{center}
\leavevmode
\epsfxsize=7.0in
\epsfysize=8.0in
   \includegraphics{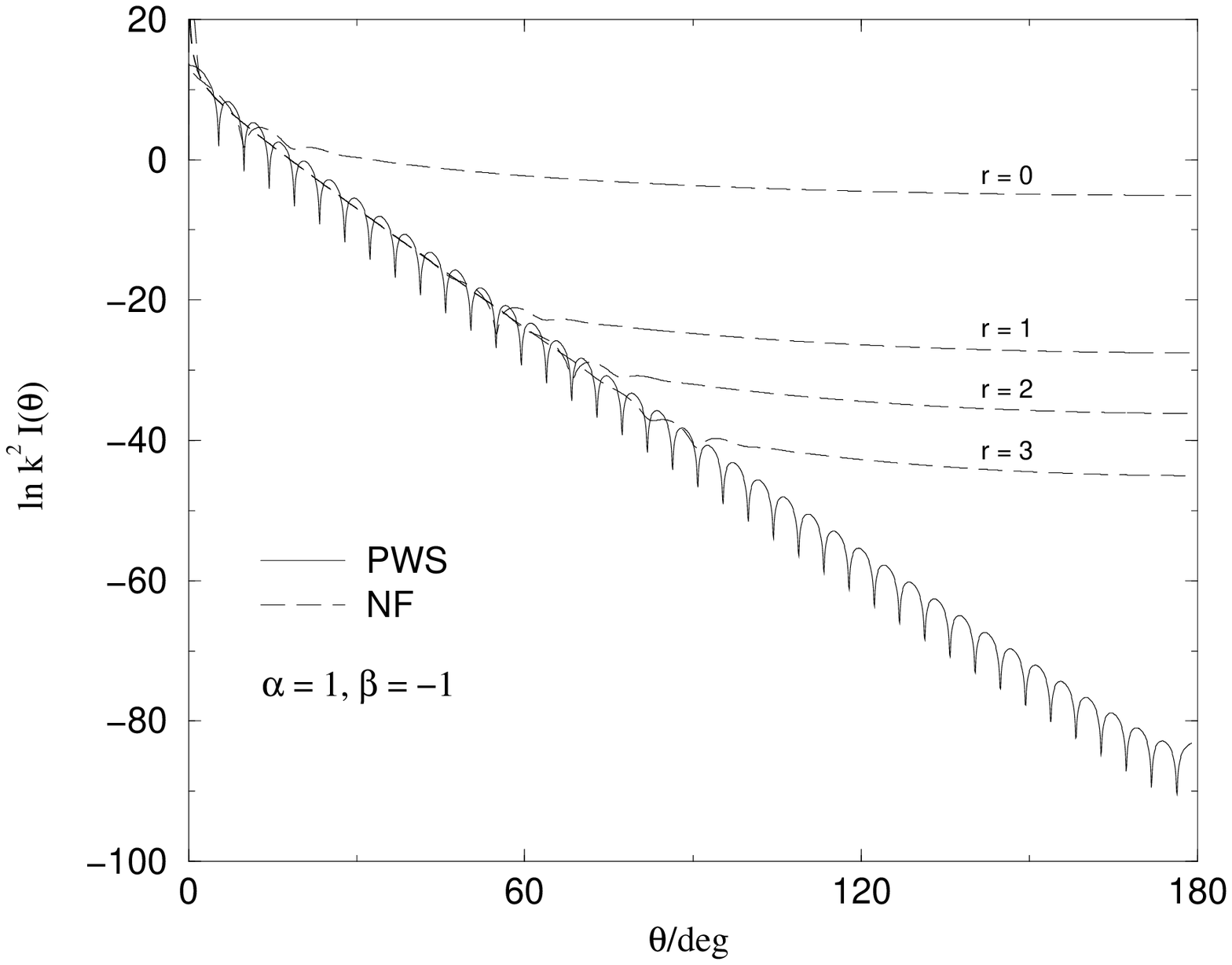}
%\epsffile{figs/295dcsop.eps}
\vspace{-0.7in}
   \caption{Plot of ln $[$ $k^2 \textit {I} (\theta)$ $]$ versus $\theta$ for the parameterization of eqns (2.1) and (2.11) with $\Lambda = 40$, $\Delta = 5$ and $(\alpha, \beta)= (1,-1)$. Solid line: PWS angular distribution. Dashed lines: N and F angular distributions, which are identically equal, for $r$ = 0,1,2 and 3 . The semiclassical NF angular distributions (not shown) pass through the oscillations of the PWS angular distribution. The NF angular distributions converge toward the semiclassical NF cross sections as $r$ increases.}
  \end{center}
\end{figure}

\clearpage

\begin{figure}[p]
  \begin{center}
\leavevmode
\epsfxsize=7.0in
\epsfysize=8.0in
   \includegraphics{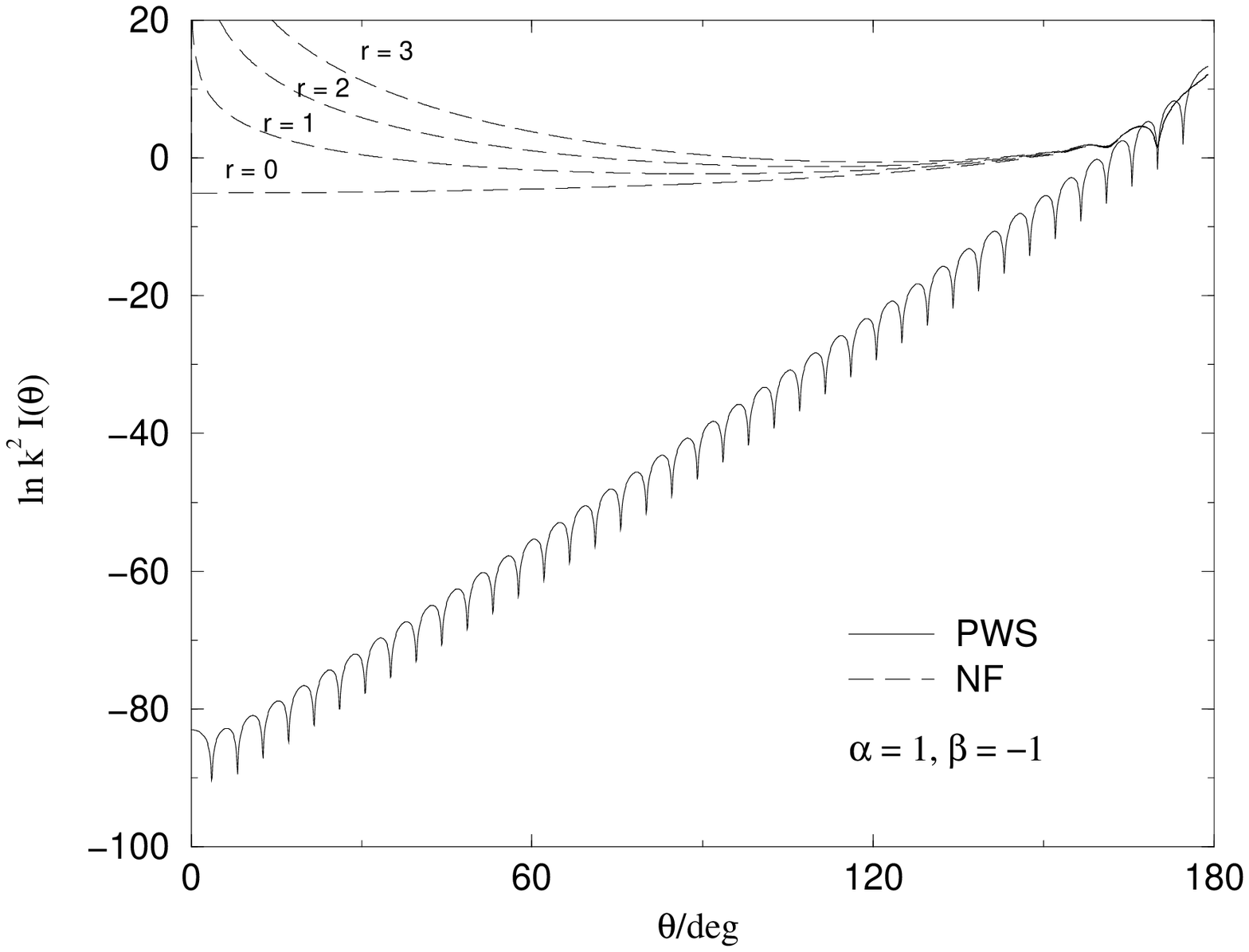}
\vspace{-0.7in}
   \caption{Plot of ln $[$ $k^2 \textit {I} (\theta)$ $]$ versus $\theta$ for the parameterization of eqns (2.1) and (2.12) with $\Lambda = 40$, $\Delta = 5$ and $(\alpha, \beta)= (1,-1)$. Solid line: PWS angular distribution. Dashed lines: N and F angular distributions, which are identically equal, for $r$ = 0,1,2 and 3 .}
  \end{center}
\end{figure}

%\clearpage
\begin{figure}[p]
  \begin{center}
\leavevmode
\epsfxsize=7.0in
\epsfysize=8.0in
   \includegraphics{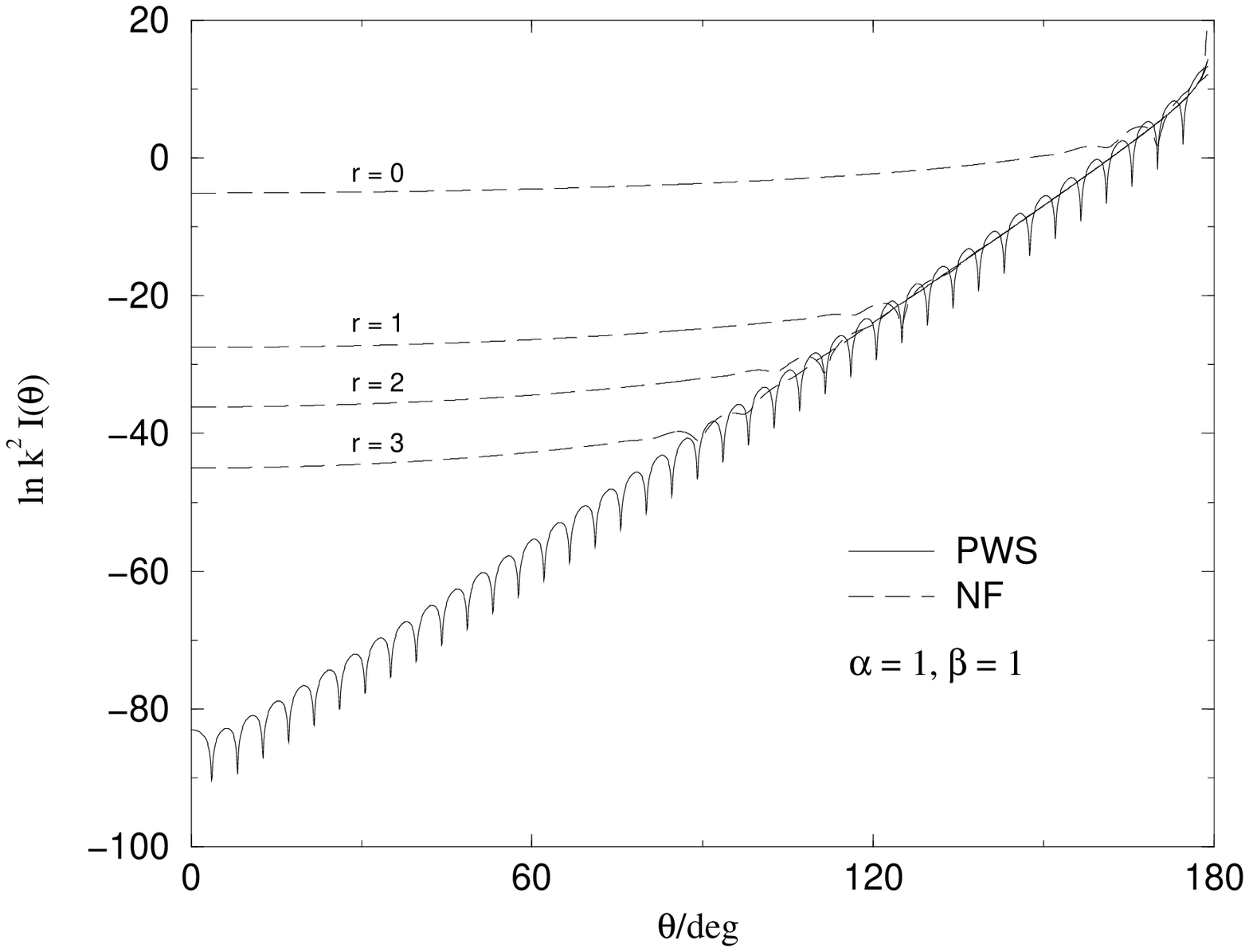}
%\epsffile{figs/213dcsop.eps}
\vspace{-0.7in}
   \caption{Plot of ln $[$ $k^2 \textit {I} (\theta)$ $]$ versus $\theta$ for the parameterization of eqns (2.1) and (2.12) with $\Lambda = 40$, $\Delta = 5$ and $(\alpha, \beta)= (1,1)$. Solid line: PWS angular distribution. Dashed lines: N and F angular distributions, which are identically equal, for $r$ = 0,1,2 and 3 .}
  \end{center}
\end{figure}

%\clearpage
\begin{figure}[p]
  \begin{center}
\leavevmode
\epsfxsize=7.0in
\epsfysize=8.0in
   \includegraphics{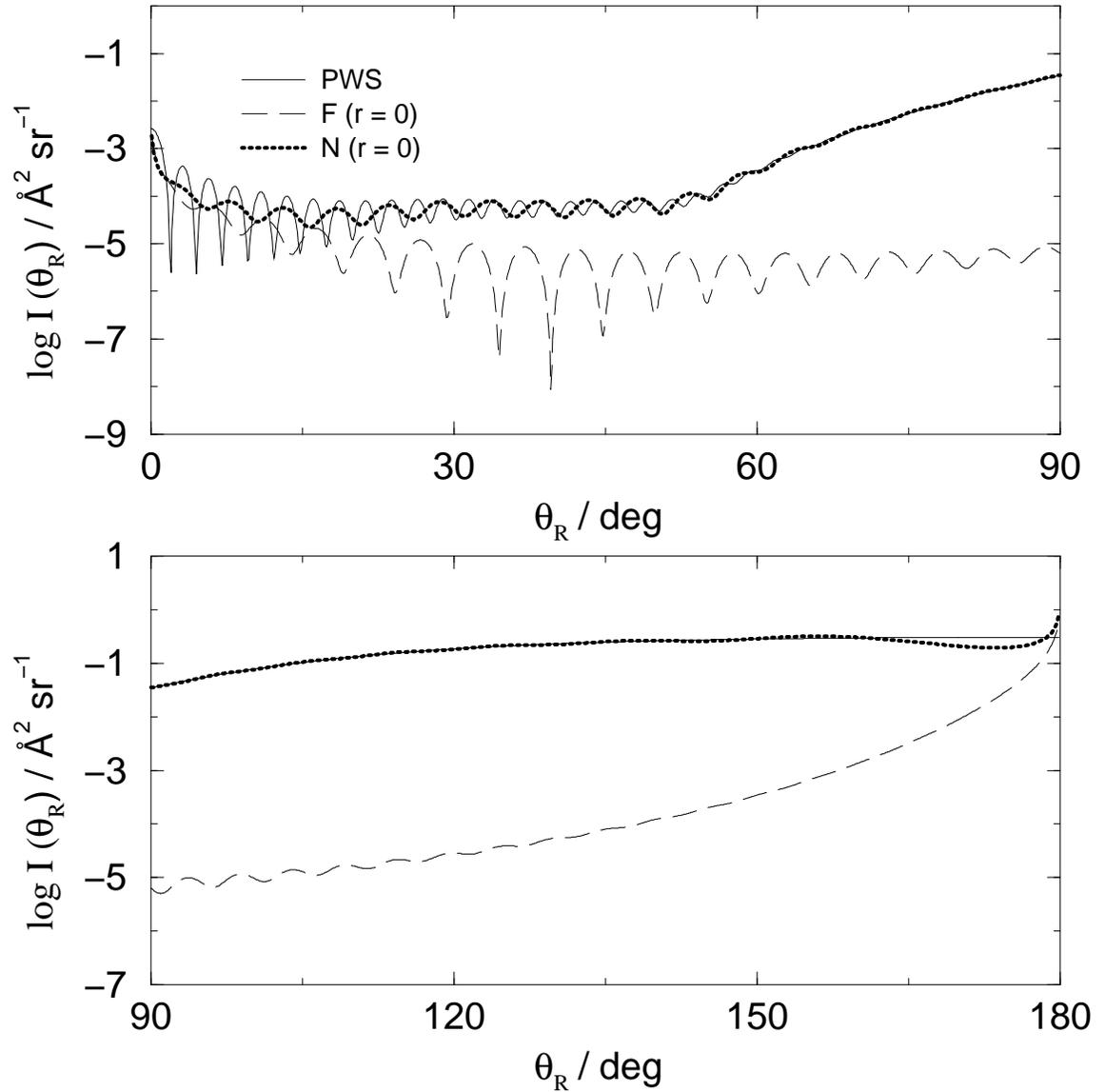}
%\epsffile{figs/295rp.eps}
\vspace{-0.4in}
   \caption{Plot of log ${I} (\theta_R)$ versus $\theta_R$ for the $\textrm {I + HI}(v_i = 0, j_i = 4) \rightarrow \textrm {IH}(v_f = 0, j_f = 4)  + \textrm {I}$ reaction at $E$ = 29.5 meV. Solid line: PWS angular distribution. Dotted line: N angular distribution for $r$ = 0. Dashed line: F angular distribution for $r$ = 0.} 
  \end{center}
\end{figure}

\clearpage
\begin{figure}[p]
  \begin{center}
\leavevmode
\epsfxsize=7.0in
\epsfysize=8.0in
   \includegraphics{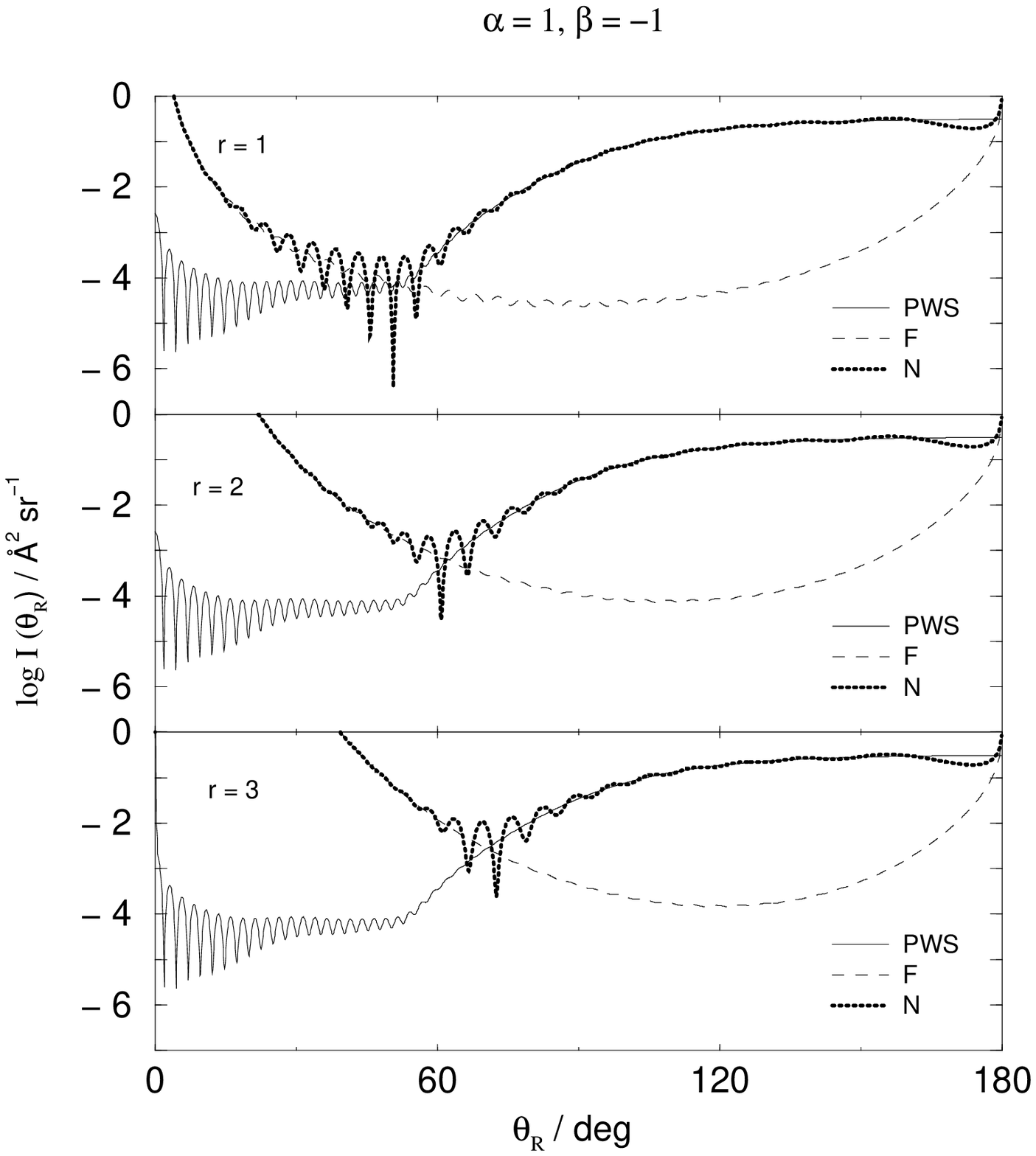}
%%\epsffile{figs/268rp.eps}
\vspace{-0.4in}
   \caption{Plot of log ${I} (\theta_R)$ versus $\theta_R$ for the $\textrm {I + HI}(v_i = 0, j_i = 4) \rightarrow \textrm {IH}(v_f = 0, j_f = 4) + \textrm {I}$ reaction at $E$ = 29.5 meV. Solid line: PWS angular distribution. Dotted line: N angular distribution for $r$ = 1,2 and 3 and  $(\alpha, \beta)= (1,-1)$. Dashed lines: F angular distribution for $r$ = 1,2 and 3 and  $(\alpha, \beta)= (1,-1)$.}
  \end{center}
\end{figure}

\clearpage
\begin{figure}[p]
  \begin{center}
\leavevmode
\epsfxsize=7.0in
\epsfysize=8.0in
   \includegraphics{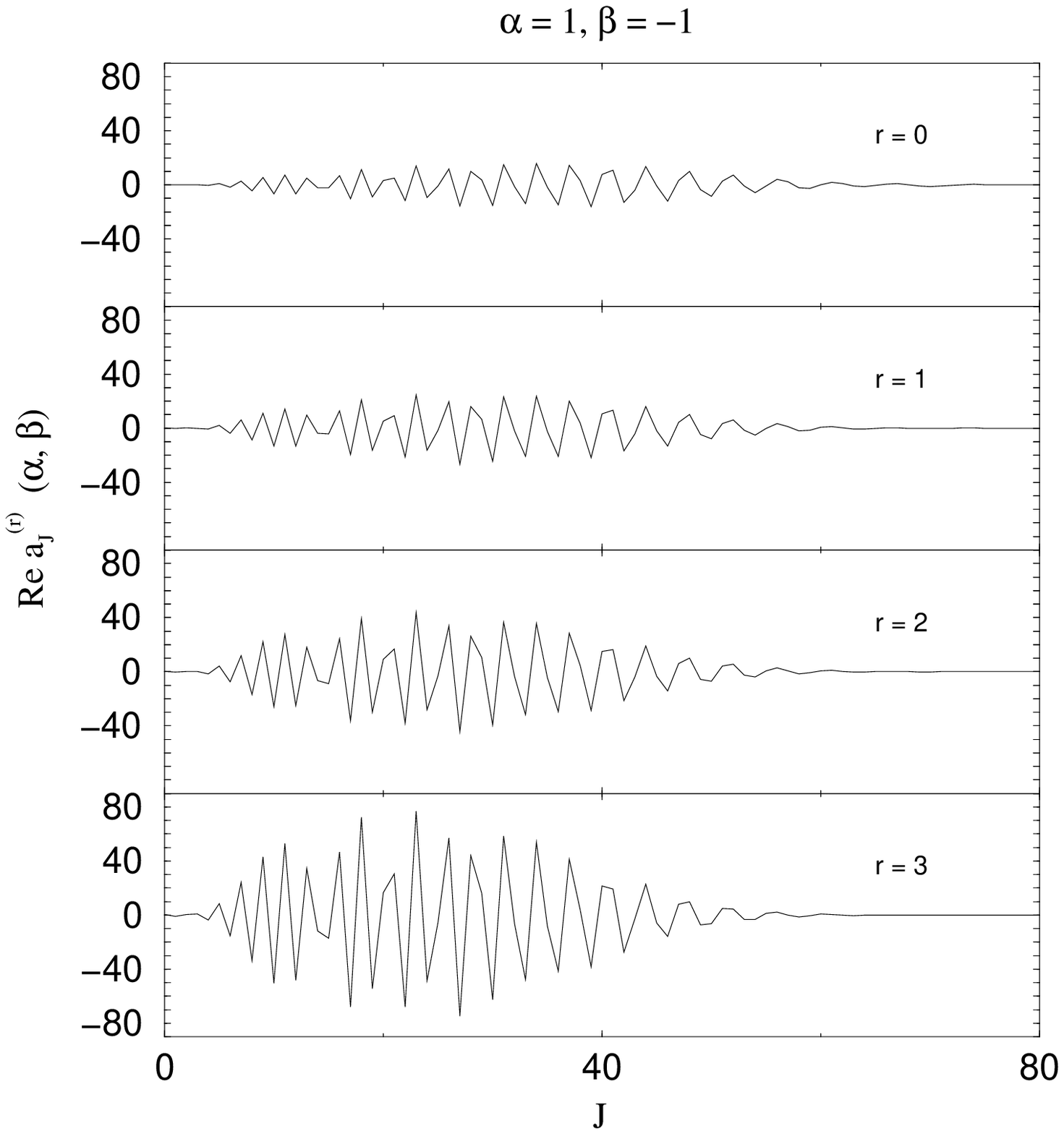}
%%\epsffile{figs/213rp.eps}
\vspace{-0.5in}
   \caption{Plot of Re $a_J^{(r)}(1,-1)$ versus $J$ for the $\textrm {I + HI}(v_i = 0, j_i = 4) \rightarrow \textrm {IH}(v_f = 0, j_f = 4) + \textrm {I}$ reaction at $E$ = 29.5 meV for $r$ = 0,1,2 and 3. $[$ $a_J^{(r)}(\alpha, \beta)$ is independent of $\alpha, \beta$ for $r$ = 0 $]$.}
  \end{center}
\end{figure}

%\clearpage
\begin{figure}[p]
  \begin{center}
\leavevmode
\epsfxsize=7.0in
\epsfysize=8.0in
   \includegraphics{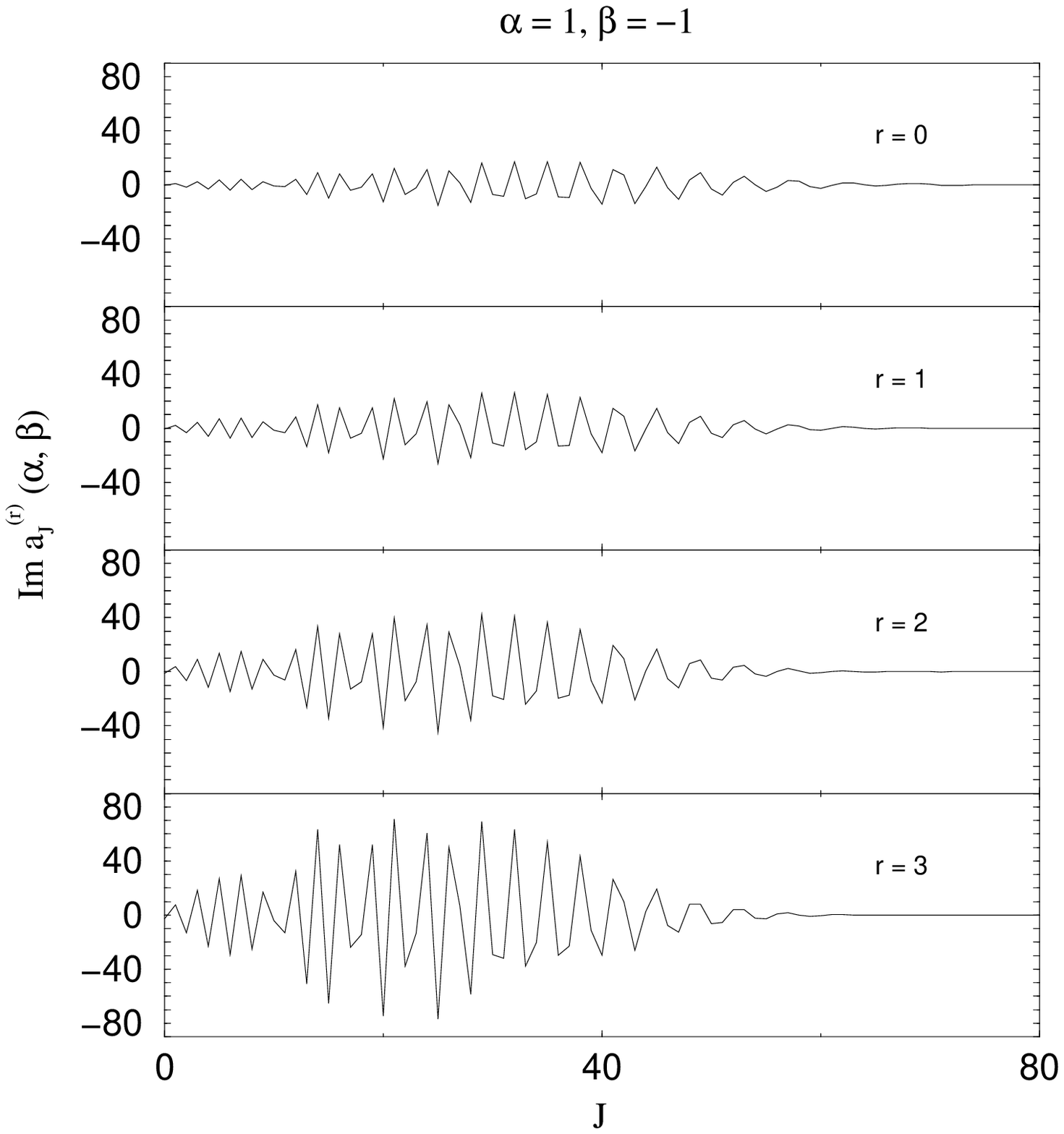}
%\epsffile{figs/295nf.eps}
\vspace{-0.5in}
   \caption{Plot of Im $a_J^{(r)}(1,-1)$ versus $J$ for the $\textrm {I + HI}(v_i = 0, j_i = 4) \rightarrow \textrm {IH}(v_f = 0, j_f = 4) + \textrm {I}$ reaction at $E$ = 29.5 meV for $r$ = 0,1,2 and 3. $[$ $a_J^{(r)}(\alpha, \beta)$ is independent of $\alpha, \beta$ for $r$ = 0 $]$.}
  \end{center}
\end{figure}

%\clearpage
\begin{figure}[p]
  \begin{center}
\leavevmode
\epsfxsize=7.0in
\epsfysize=8.0in
   \includegraphics{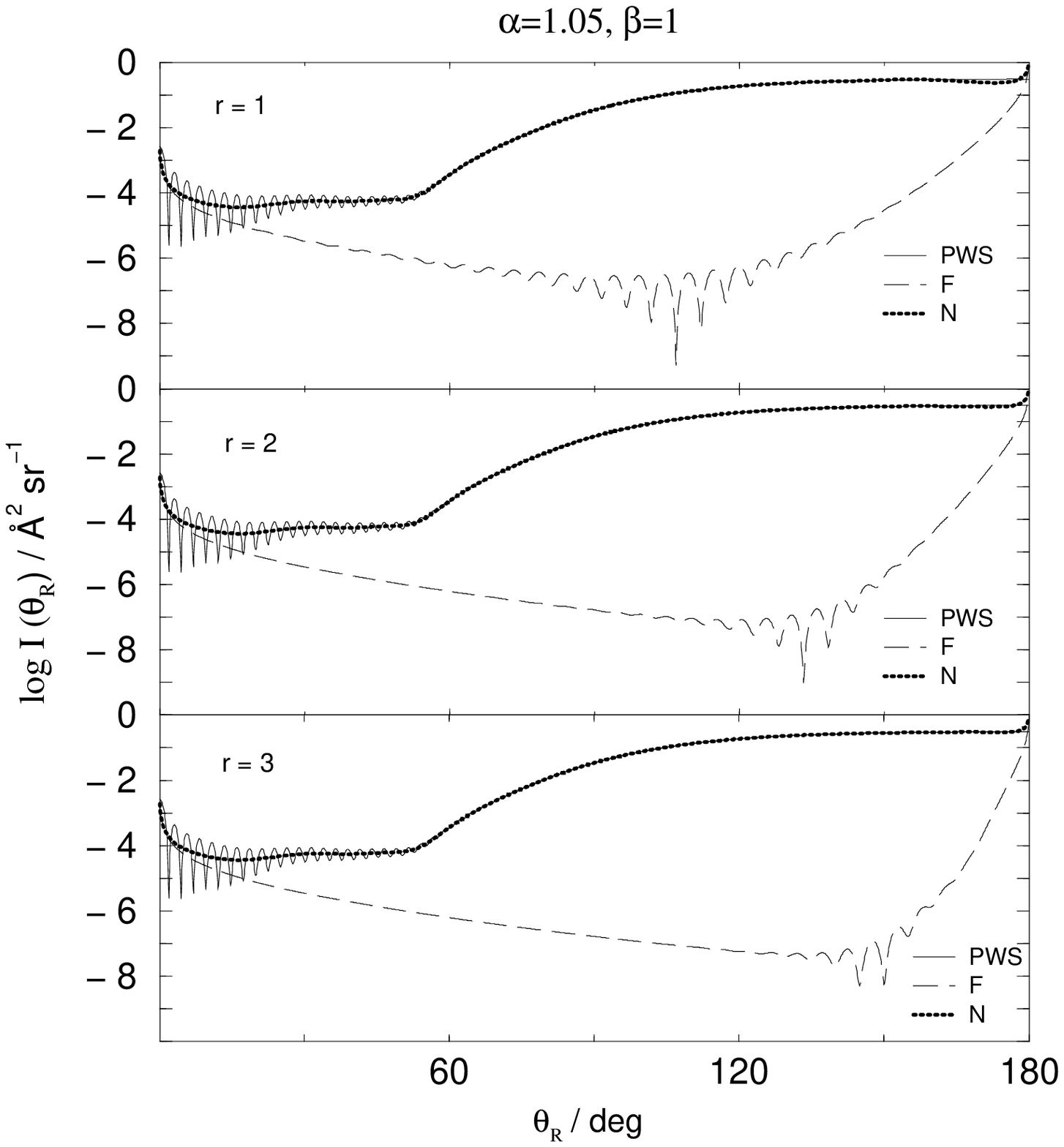}
%%\epsffile{figs/268nf.eps}
\vspace{-0.4in}
   \caption{Plot of log ${I} (\theta_R)$ versus $\theta_R$ for the $\textrm {I + HI}(v_i = 0, j_i = 4) \rightarrow \textrm {IH}(v_f = 0, j_f = 4) + \textrm {I}$ reaction at $E$ = 29.5 meV. Solid line: PWS angular distribution. Dotted line: N angular distribution for $r$ = 1,2 and 3 and  $(\alpha, \beta)= (1.05,1)$. Dashed lines: F angular distribution for $r$ = 1,2 and 3 and  $(\alpha, \beta)= (1.05,1)$.}
  \end{center}
\end{figure}

%\clearpage
\begin{figure}[p]
  \begin{center}
\leavevmode
\epsfxsize=7.0in
\epsfysize=8.0in
   \includegraphics{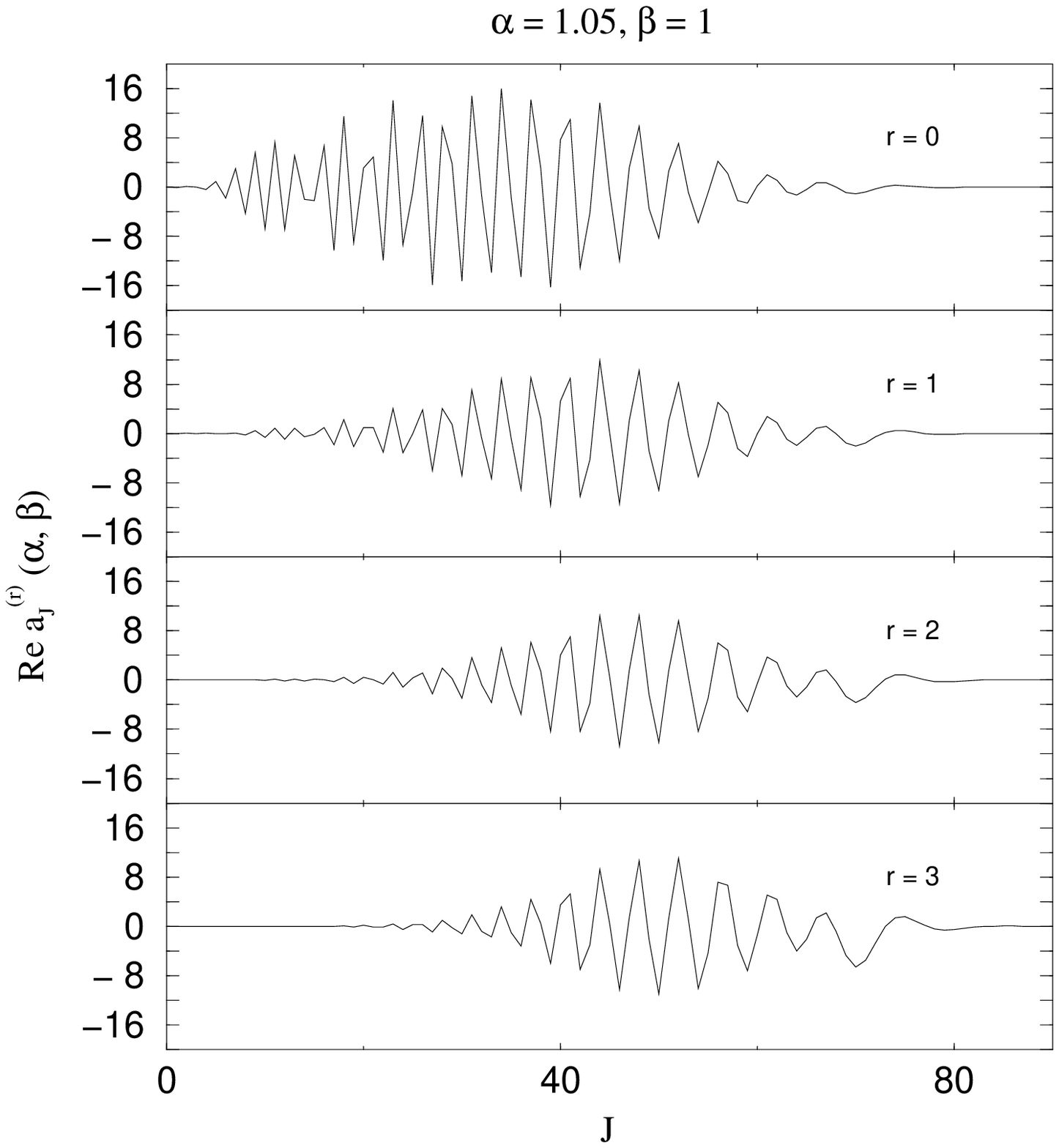}
%%\epsffile{figs/213nf.eps}
\vspace{-0.5in}
   \caption{Plot of Re $a_J^{(r)}(1.05,1)$ versus $J$ for the $\textrm {I + HI}(v_i = 0, j_i = 4) \rightarrow \textrm {IH}(v_f = 0, j_f = 4) + \textrm {I}$ reaction at $E$ = 29.5 meV for $r$ = 0,1,2 and 3. $[$ $a_J^{(r)}(\alpha, \beta)$ is independent of $\alpha, \beta$ for $r$ = 0 $]$.}
  \end{center}
\end{figure}

%\clearpage
\begin{figure}[p]
  \begin{center}
\leavevmode
\epsfxsize=7.0in
\epsfysize=8.0in
   \includegraphics{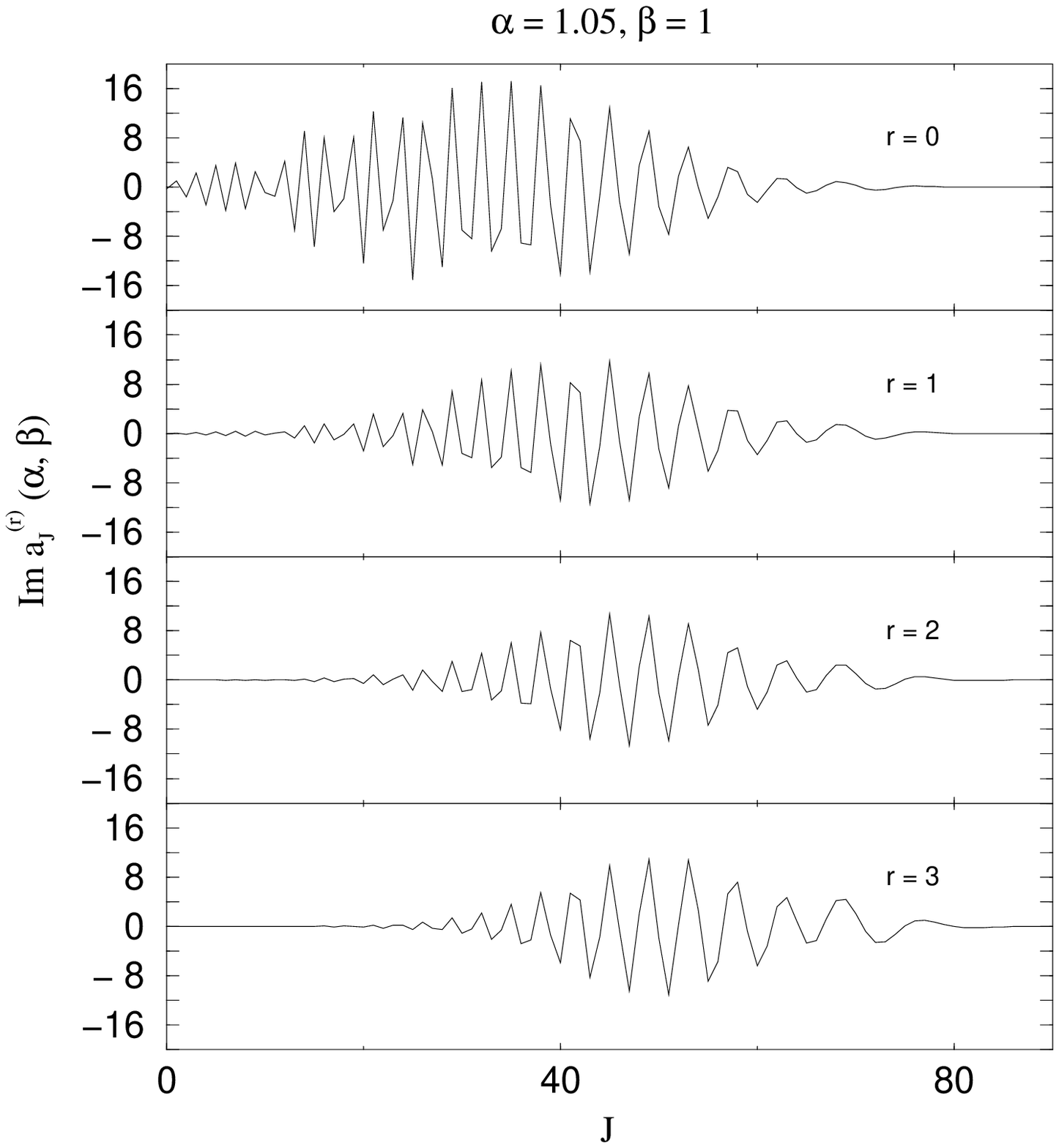}
\vspace{-0.5in}
   \caption{Plot of Im $a_J^{(r)}(1.05,1)$ versus $J$ for the $\textrm {I + HI}(v_i = 0, j_i = 4) \rightarrow \textrm {IH}(v_f = 0, j_f = 4) + \textrm {I}$ reaction at $E$ = 29.5 meV for $r$ = 0,1,2 and 3. $[$ $a_J^{(r)}(\alpha, \beta)$ is independent of $\alpha, \beta$ for $r$ = 0 $]$.}
  \end{center}
\end{figure} 
\end{document}